\documentclass[aps,floats,epsf,twocolumn]{revtex4}
 \usepackage{graphics}

\begin{document}

\title{Interactions and phase transitions on graphene's honeycomb lattice}

\author{ Igor F. Herbut}

\affiliation{ Department of Physics, Simon Fraser University, 
 Burnaby, British Columbia, Canada V5A 1S6}

\begin{abstract} The low-energy theory of interacting electrons
on graphene's two-dimensional honeycomb lattice is derived and
discussed. In particular, the Hubbard model in the large-$N$ limit
is shown to have a semi-metal - antiferromagnetic insulator quantum
critical point in the universality class of the Gross-Neveu model.
The same equivalence is conjectured to hold in the physical case
$N=2$, and its consequences for various physical quantities are
examined. The effects of the long-range Coulomb interaction and of
the magnetic field are discussed.
\end{abstract}
\maketitle

\vspace{10pt}

A graphite monolayer, or {\it graphene}, emerged recently as the new
frontier in physics of electronic systems with reduced
dimensionality \cite{novoselov}. Such two-dimensional, or
quasi-two-dimensional systems have led to some of the most startling
discoveries in the condensed matter physics in the recent past, the
quantum Hall effects and the  metal-insulator transitions in
silicon-MOSFETS and Ga-As heterostructures, and the high-temperature
superconductivity in cuprates being prime examples. What makes
graphene qualitatively new is its semi-metallic nature with
low-energy quasiparticles behaving as `relativistic' Dirac spinors
over a good portion of the conducting band. The spinor structure is
a general consequence of the bipartite nature of the honeycomb
lattice \cite{semenoff}. Indeed, recently observed quantization
rules for the  Hall conductivity \cite{novoselov1} may be understood
as a direct consequence of the Dirac nature of its low-energy
spectrum \cite{gusynin}, \cite{peres}.

The relativistic spectrum and the concomitant linearly vanishing
density of states at the Fermi level, similarly as in the
superconducting state of cuprates, provide graphene's quasiparticles
with an additional protection against the effects of interactions.
Nevertheless, a sufficiently strong repulsion is expected to turn
the semi-metallic state into a gapped insulator, possibly breaking
the translational and/or the rotational symmetry in the process.
Within the simplest interacting theory defined by the Hubbard model
there is convincing numerical evidence for the quantum phase
transition at a large Hubbard $U$ into an antiferromagnet (AF)
\cite{sorella}. On the other hand, the long-range Coulomb
interaction remains unscreened in the semi-metal (SM)
\cite{gonzales}, and has been argued to favor the
charge-density-wave (CDW) at strong coupling \cite{khveshchenko}.
The competition between different instabilities, the universality
class, or even the order, of the SM - insulator transition, and the
interplay of interactions with the Landau quantization in the
external magnetic field present some of the basic open problems.
Although graphene in its natural state may not be near a critical
point \cite{wagner}, one can conceive mechanical deformations that
would pull it deeper into the strong-coupling regime \cite{gloor}.
Finally, the outcome of the competition between different
interactions should have consequences for the selection of the
ground state in the magnetic field, even at weak coupling
\cite{miransky}.

In the present communication some of these issues are addressed by
considering the half-filled Hubbard model on a honeycomb lattice,
complemented with the additional long-range Coulomb interaction
between electrons. The analysis is based on a useful decomposition
of Hubbard's on-site interaction on a bipartite lattice into a sum
of squares of average and staggered densities, and average and
staggered magnetizations. The long-range part of the Coulomb
interaction may be represented by a massless scalar gauge field,
whereas its main effect on the lattice scale is to provide the
repulsion between nearest-neighbors. When prepared like this, in the
continuum limit such an extended Hubbard model on a honeycomb
lattice maps onto a $2+1$-dimensional field theory of Dirac
fermions, with {\it nine} different couplings. Its apparent
complexity notwithstanding, when generalized to a large number of
fermion flavors $N$, the theory admits a simple SM - AF  critical
point of the Gross-Neveu variety \cite{gross}. Coulomb interaction
is marginally irrelevant at the critical point. Assuming that the
equivalence with the Gross-Neveu model persists down to the physical
case of $N=2$, I infer the values of the critical exponents in the
original Hubbard model. A more general phase diagram, and the
implications of these results for graphene are discussed.

The extended Hubbard model will be defined by the Hamiltonian $H=
H_0 + H_1$ where
\begin{equation}
H_0= -t \sum_{\vec{A}, i, \sigma=\pm 1} u^\dagger _\sigma (\vec{A})
v_\sigma (\vec{A}+\vec{b}_i) + H. c.,
\end{equation}
\begin{equation}
H_1 = \sum_{\vec{X}, \vec{Y}, \sigma, \sigma' } n_\sigma (\vec{X}) [
\frac{U}{2} \delta_{\vec{X},\vec{Y}} + \frac{ e^2
(1-\delta_{\vec{X},\vec{Y}}) }{ 4 \pi |\vec{X}-\vec{Y}|} ]
n_{\sigma'} (\vec{Y}).
\end{equation}
The sites $\vec{A}$ denote one triangular sublattice of the
hexagonal lattice, generated by linear combinations of the basis
vectors $\vec{a}_1= (\sqrt{3}, -1)(a/2)$, $\vec{a}_2 = (0,a)$. The
second sublattice is then at $\vec{B}=\vec{A}+\vec{b}$, with the
vector $\vec{b}$ being either $\vec{b}_1= (1/\sqrt{3},1) (a/2)$,
$\vec{b}_2= (1/\sqrt{3},-1) (a/2)$, or $\vec{b}_3= (-a/\sqrt{3},0)$.
$a$ is the lattice spacing. Neutralizing background is assumed, as
usual.

The doubly degenerate spectrum of $H_0$ at $E(\vec{k})=\pm t
|\sum_{i} \exp [\vec{k}\cdot \vec{b}_i] |$ becomes linear and
isotropic in the vicinity of two non-equivalent points at the edges
of the Brillouin zone at $\pm \vec{K}$, with $\vec{K} =
(1,1/\sqrt{3}) (2\pi/a \sqrt{3})$ \cite{semenoff}. Retaining only
the Fourier components near $\pm \vec{K}$ one can write the
quantum-mechanical action corresponding to $H_0$ at low energies as
$S=\int_0 ^{1/T} d\tau d\vec{x} L_0 $, with the {\it free}
Lagrangian $L_0$ defined as
\begin{equation}
L_0= \sum_{\sigma=\pm 1}
\bar{\Psi}_\sigma (\vec{x},\tau) \gamma_\mu
\partial_\mu \Psi_\sigma (\vec{x},\tau),
\end{equation}
and
\begin{eqnarray}
\Psi_\sigma ^\dagger (\vec{x},\tau) = T \sum_{\omega_n} \int^\Lambda
\frac{d\vec{q}}{(2\pi a )^2} e^{i\omega_n \tau + i \vec{q}\cdot
\vec{x}} (u^\dagger _\sigma (\vec{K}+\vec{q},\omega_n), \\ \nonumber
 v^\dagger _\sigma (\vec{K}+\vec{q},\omega_n),
 u^\dagger _\sigma(-\vec{K}+\vec{q},\omega_n), v^\dagger _\sigma
(-\vec{K}+\vec{q},\omega_n) ),
\end{eqnarray}
where it was convenient to rotate the reference frame so that $q_x =
\vec{q}\cdot \vec{K}/K$ and $q_y = (\vec{K}\times \vec{q})\times
\vec{K}/K^2 $, and set $\hbar=k_B=v_F =1$, where $v_F =ta\sqrt{3}/2$
is the Fermi velocity. Choosing $\gamma_0= I_2 \otimes \sigma_z$
implies $\gamma_1 = \sigma_z \otimes \sigma_y$ and $\gamma_2 = I_2
\otimes \sigma_x$, with $I_2$ as the $2 \times 2$ unit matrix, and
$\vec{\sigma}$ as the Pauli matrices. $\Lambda\approx 1/a $ is the
ultraviolet cutoff over which the linear approximation for the
dispersion holds. The summation convention is adopted hereafter, but
{\it only} over repeated space-time indices. Besides the
`relativistic' invariance, $L_0$ also exhibits a global invariance
under the $U(4)$, generated by $\{I_2, \vec{\sigma} \} \otimes \{ I,
\gamma_3,\gamma_5, \gamma_{35} \}$, where $I$ is the $4 \times 4$
unit matrix, $\gamma_3= \sigma_x \otimes \sigma_y$, $\gamma_5 =
\sigma_y \otimes \sigma_y$, and $\gamma_{35}= i \gamma_3 \gamma_5$.
This is similar to the emergent `chiral' symmetry of a d-wave
superconductor \cite{herbut}.

Generalizing slightly Hamman's decomposition, the first term in
$H_1$ can also be rewritten exactly as
\begin{eqnarray}
\frac{U}{8} \sum_{\vec{A}} [ (n(\vec{A} ) + n(\vec{A}+\vec{b}) )^2 +
(n(\vec{A} ) - n(\vec{A}+\vec{b}) )^2 \\ \nonumber -(m(\vec{A} ) +
m(\vec{A}+\vec{b}) )^2 - (m(\vec{A} ) - m(\vec{A}+\vec{b}) )^2 ]
\end{eqnarray}
where $n(\vec{A}),m(\vec{A})  = u^\dagger_+ (\vec{A}) u_+ (\vec{A})
\pm  u^\dagger_- (\vec{A}) u_- (\vec{A}) $, are the particle number
and the magnetization at the sites $\vec{A}$. Variables at the
second sublattice are analogously defined in terms of
$v_\sigma(\vec{B}) $.

Defining the two {\it slow} components of the fields as
\begin{equation}
r_\sigma  ^{1,2} (\vec{x},\tau) = \int_{|\vec{k}\pm \vec{K}|<
\Lambda} \frac{d\vec{k}}{(2\pi)^2} e^{i \vec{k}\cdot \vec{x} }
r_{\sigma} (\vec{k}, \tau),
\end{equation}
with $r=u,v$, the Dirac field becomes
\begin{eqnarray}
\Psi^\dagger_\sigma (\vec{x},\tau) e^{i (\vec{K}\cdot \vec{x})
\gamma_{35} }= (u^{1 \dagger} _\sigma(\vec{x},\tau),  \\ \nonumber
v^{1 \dagger}_\sigma(\vec{x},\tau),  u^{2 \dagger} _\sigma
(\vec{x},\tau),  v^{2 \dagger} _\sigma (\vec{x},\tau)).
\end{eqnarray}
At low energies one may then approximate
\begin{equation}
r_\sigma (\vec{x},\tau)\approx
 r_\sigma ^{1} (\vec{x},\tau)+ r_\sigma ^{2} (\vec{x},\tau),
\end{equation}
so that the spin densities on the two sublattices become
\begin{eqnarray}
r_\sigma ^{\dagger}(\vec{x},\tau) r_\sigma (\vec{x},\tau)\approx
\\ \nonumber
 \frac{1}{2} \bar{\Psi}_\sigma  (\vec{x},\tau) (I_2 + e^{i
2\vec{K}\cdot \vec{x} \sigma_z} \sigma_x ) \otimes (  \sigma_z \pm
I_2 ) \Psi _\sigma (\vec{x},\tau),
\end{eqnarray}
with the plus sign for $r=u$ and the minus for $r=v$.

\begin{figure}[t]
{\centering\resizebox*{80mm}{!}{\includegraphics{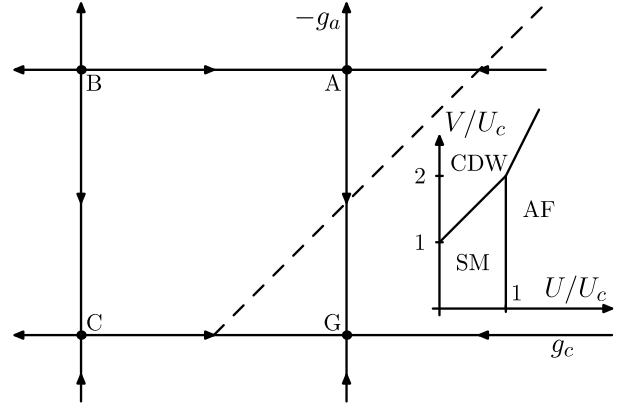}}
\par} \caption[] {The large-$N$ flow diagram in the attractive plane
$e=g_d=g_f=\tilde{g}_x =0$. A, B, C, and G are the AF, bicritical,
CDW, and the Gaussian fixed points, respectively. A and C are in the
Gross-Neveu universality class. The extended Hubbard model in Eq.
(2) defines the (dashed) line of initial conditions $g_c =
(U-V)/4U_c$ and $g_a = -U /4U_c$, with $V$ as a fixed
nearest-neighbor repulsion. Inset: the resulting phase diagram. }
\end{figure}

The notation is now in place to write the low-energy theory of the
extended Hubbard model. In the continuum limit ($a\rightarrow 0$)
the Lagrangian becomes
\begin{equation}
L= L_0 -i a_0 \sum_{\sigma} \bar{\Psi}_\sigma \gamma_0 \Psi_\sigma +
a_0 \frac{|\nabla|}{2 e^2}  a_0 + \sum_{x=d,c,f,a} L_x,
\end{equation}
with
\begin{eqnarray}
L_x = g_x (\sum_{\sigma } w_{x,\sigma} \bar{\Psi}_\sigma M_x
\Psi_\sigma)^2 + \\ \nonumber \tilde{g}_x \sum_{\mu=3,5}
(\sum_{\sigma \pm 1} w_{x,\sigma} \bar{\Psi}_\sigma M_x \gamma_1
\gamma_\mu \Psi_\sigma)^2,
\end{eqnarray}
and $w_{d,\sigma}=w_{c,\sigma}=1$,
$w_{f,\sigma}=w_{a,\sigma}=\sigma$, $M_d = M_f = \gamma_0$ and $M_c
= M_a = I$. The short-range couplings are $g_d= -2\tilde{g}_d - e^2
/4K=(U+V) a^2 /8$, $g_c=-2 \tilde{g}_c =  (U-V) a^2 /8$, $g_f=g_a =
-2 \tilde{g}_f = -2\tilde{g}_a = -Ua^2 /8$. $d$ and $c$ couplings
correspond to the first (average density) and the second (staggered
density), whereas $f$ and $a$ couplings represent the third
(magnetization) and the fourth (staggered magnetization) terms in
Eq. (5). The Coulomb interaction is represented by: 1) the
intra-unit-cell, nearest-neighbor repulsion $V= e^2 \sqrt{3} /
(a\pi)$, and  the $2\vec{K}$ Fourier component $e^2 / 2K$, and 2)
its long-range part, which is recovered upon Gaussian integration
over the scalar gauge-field $a_0$. $|\nabla|$ should be understood
as $|\vec{q}|$ in Fourier space \cite{herbut1}. Whereas such a
separation would be exact for an infinitely long-ranged interaction,
it is only an {\it approximation} for the Coulomb interaction.

The usual power counting implies that all short-range interactions
in $L$ are irrelevant, and that the charge $e$ is a marginal
coupling at the non-interacting fixed point $g_x = \tilde{g}_x = e
=0$. Any critical point would therefore have to lie in the
 strong-coupling regime. To exert some control over it
 we may deform the Lagrangian from two to $N$ flavors of the Dirac
 fields as follows:
 \begin{equation}
 \bar{\Psi}_+ \Psi_+  \rightarrow
 \sum_{\sigma=1}^{N/2} \bar{\Psi}_\sigma \Psi_\sigma;
 \bar{\Psi}_- \Psi_-  \rightarrow
 \sum_{\sigma=(N/2) +1 }^{N} \bar{\Psi}_\sigma \Psi_\sigma,
 \end{equation}
and $g_x\rightarrow 2 g_x/N$, $\tilde{g}_x \rightarrow 2
\tilde{g}_x/N$, $e^2 \rightarrow 2 e^2 /N$. The integration over the
Fourier components with $\Lambda/b < q <\Lambda$ and $-\infty <
\omega < \infty $ renormalizes then the short-range couplings at
$T=0$ as
\begin{equation}
\beta_{x}= \frac{dg_{x}}{d\ln b} = - g_{x} -  C_x g_{x} ^2 + O(1/N),
\end{equation}
\begin{equation}
\tilde{\beta}_x = \frac{d\tilde{g}_x} {d\ln b} = - \tilde{g}_x + 2
\tilde{g}_x ^2 + O(1/N),
\end{equation}
with $C_{c,a}=4$, $C_{d,f}=0$, and with the couplings rescaled as
$g\Lambda/\pi \rightarrow g$. To the leading order in $1/N$
$\beta$-functions for different interactions thus do not mix
\cite{kaveh}. Since the model when $N=\infty$ is exactly solvable by
the saddle-point method, the leading order $\beta$-functions may
also be understood as guaranteing that the solution is cutoff
independent \cite{kaveh}. Non-analyticity of the inverse gauge-field
propagator and the gauge invariance of $L$ also dictate that
\begin{equation}
\beta_e = \frac{d e^2}{d\ln b} = (z-1) e^2,
\end{equation}
with $z$ as the dynamical exponent, {\it exactly} \cite{herbut1}.
Relativistic invariance of $L$ is broken when $e\neq 0$, and
consequently $z\neq 1$ at finite length scales. Similarly to the
bosonic case,
\begin{equation}
z= 1-\frac{e^2}{2\pi N} +O(1/N^2),
\end{equation}
and the charge is marginally {\it irrelevant} to the order in $1/N$
\cite{gonzales}, \cite{vafek}.

   Besides the trivial fully attractive fixed point, the large-$N$
$\beta$-functions in Eqs. (13)-(15) exhibit {\it two critical
points} in the attractive plane $e^2=g_f =g_d =\tilde{g}_x=0$: 1) at
$g_a=-1/4$, $g_c=0$ and 2) $g_c=-1/4$, $g_a=0$. There is also a
bicritical point at $g_a = g_c=-1/4$, which directs the flows
towards one of the two critical points (Fig. 1). The critical points
are related by the symmetry under a change of sign of $\gamma_\mu$
for `down' components with $\sigma = N/2 +1, ... N$ accompanied by
the exchange of $g_a$ and $g_c$. The transition is either to $A=
\langle \sum_\sigma \sigma \bar{\Psi}_\sigma \Psi_\sigma \rangle\neq
0$, which corresponds to an AF with a finite staggered
magnetization, or to a CDW,  with the finite staggered density $C=
\langle \sum_\sigma \bar{\Psi}_\sigma \Psi_\sigma \rangle\neq 0 $.
The same, of course, follows from the explicit solution of the model
at $N=\infty$. The flow of $g_f$ towards the origin also agrees with
the saddle-point equations, which do not show a ferromagnetic
critical point at $N=\infty$, whereas the irrelevance of $g_d$
simply means that the chemical potential vanishes.

  Eqs. (14), however, appear to exhibit additional critical points at
  $\tilde{g}_x=1/2$. These, however, would
  occur {\it within} the AF or the CDW, and are
  artifacts of our procedure which checks only the stability of the
  semi-metal. It is easy to see from the explicit solution that
  the existing gap prevents such an additional transition. All
  $\tilde{g}_x $ are therefore irrelevant.

The transition in the pure, $e=0$, repulsive Hubbard model with $g_a
<0$ and $g_c>0$ in the large-$N$ limit is controlled therefore by
the critical point A. Recalling that $g_a = Ua^2 \Lambda/(8\pi)$,
with $\Lambda\approx 1/a$, and $t a \sqrt{3}/2 =1$ by our
convention, one finds that this corresponds to the critical value of
$U_c/t\approx 5.5$, certainly fortuitously close to the values found
in numerical calculations \cite{sorella}. Above the critical
interaction the system develops a gap, at the same time becoming
insulating and antiferromagnetic. At the critical line $g_c =
g_f=g_d= \tilde{g}_x= e^2 =0$, upon the change of sign of
$\gamma$-matrices for `down' components the Lagrangian becomes
identical to the much studied Gross-Neveu model in 2+1 dimensions
\cite{gross}. Evidently, the Gross-Neveu critical point has only one
unstable direction to the order $1/N$. Since the actual expansion
parameter is $4N$, I expect this feature to survive even for $N=2$.
This leads to the conjecture that the SM - AF
 transition in the Hubbard model is continuous and described by the
$N=2$ Gross-Neveu critical point, at which
\begin{equation}
\langle \bar{\Psi}_\sigma (\vec{q},\omega) \Psi_\sigma
(\vec{q},\omega) \rangle \sim (q^2 +\omega^2)^{(\eta_\Psi - 1 )/2},
\end{equation}
with the fermion's anomalous dimension $\eta_\Psi = (2/(3 \pi^2 N))
+ O(1/N^2)$ \cite{vasilev}. The order parameter's correlation
function  at the critical point also decays as:
\begin{equation}
\langle A(\vec{x},\tau) A(0,0) \rangle \sim ( x^2 + \tau^2
)^{-(1+\eta)/2},
\end{equation}
where $\eta$ is the standard anomalous dimension, and $\eta = 1-
16/(3\pi^2 N) +O(1/N^2)$. The correlation length diverges at the
critical point with the exponent $\nu = 1+ 8/(3\pi^2 N) + O(1/N^2)$,
and the usual scaling laws are expected to be satisfied. The
critical exponents have been computed to the order $1/N^2$ (with
$\eta_\Psi$ known even to the order $1/N^3$) \cite{vasilev}, as well
as being determined by Monte Carlo calculations, the
$\epsilon$-expansion \cite{karkkainen}, and the exact
renormalization group \cite{wetterich}. In summary, for $N=2$ one
finds $\eta_\Psi = 0.038 \pm 0.006$, $\nu= 0.97 \pm 0.07$, and $\eta
= 0.770 \pm 0.016$ \cite{wetterich}.

The presence of gapless fermions on the semi-metallic side places
the Gross-Neveu phase transition outside the usual
Ginzburg-Landau-Wilson paradigm, as evidenced by the large anomalous
dimension $\eta$, for example. In fact, the Gross-Neveu model
probably defines the simplest such a universality class. Its
distinct characteristic is the fermion's anomalous dimension
$\eta_\Psi$, which governs the disappearance of quasiparticles as
the transition is approached on the semi-metallic side. Scaling
dictates that the residue of the quasiparticle pole behaves as
\begin{equation}
Z_\Psi \sim (U_c - U)^{\eta_\Psi},
\end{equation}
so that a very small $\eta_\Psi$ would make it appear discontinuous
at $U=U_c$.

  In the AF, the eight generators that anticommute with
  $\sigma_z \otimes \gamma_0$ become broken; these are $(I_2,
  \sigma_z) \otimes (\gamma_3, \gamma_5)$ and $(\sigma_x, \sigma_y)
  \otimes (I, \gamma_{35})$. Among these only
  $(\sigma_x,\sigma_y)\otimes I$, which
  generate the usual spin rotations,
  correspond to the exact symmetry at $U=0$, whereas
  the rest emerge as generators of approximate symmetries only at low energies. In
  the insulating phase the Goldstone bosons which correspond to the
  emerging generators are gapped, due to the irrelevant
  terms excluded from $L$ \cite{seradjeh}. The low-energy spectrum in
  the insulator consists therefore only of the usual magnons.

The long-range nature of Coulomb interaction is found to be
irrelevant at a large $N$. On the scale of lattice spacing, however,
Coulomb interaction leaves its imprint on the initial value of the
coupling $g_c$, as indicated right below Eq. (11), for example
\cite{alicea}. In general, if the nearest-neighbor repulsion $V$ is
made sufficiently strong so that the line of initial conditions in
Fig. 1 reaches left of the point C, there is an additional
semi-metal - CDW transition. Identifying $V$ with $\sim e^2/a$ gives
an alternative mechanism to that of ref. 8 for the CDW formation.
The two lines of continuous transitions merge above a certain $V$,
when the line of initial conditions comes left of the point B. The
direct transition between the AF and the CDW is discontinuous. It
seems natural, however, to assume that in reality $U>V$, which would
suggest a single, continuous, antiferromagnetic transition.

  For graphene, $t\approx  2.5 eV $, $U\approx 5 - 12 eV$,
  $ U/V \approx 2 - 3$ \cite{gloor}, so that
the system is probably on the SM side of the transition. The
external magnetic field, however, changes the density of states into
a series of delta-functions, so that the transition can now in
principle take place even at an infinitesimal coupling
\cite{miransky}. In the magnetic field the flow of the couplings
should be cutoff at $\sim 1/l_B$, where $l_B \gg a$ is the magnetic
length. If the large-N picture presented here holds for $N=2$, in
the the pure, $e=0$, Hubbard model, at a sufficiently low field all
couplings would become negligible compared to $g_a$. This would
suggest that the magnetic field, at least with the Zeeman term
neglected and the long-range component of the interaction screened
by a metallic substrate, for example, should `catalyze' the
antiferromagnetic order at a weak $U$.

The author thanks G. Sawatzky for a helpful correspondence.
 This work has been supported by NSERC of Canada.

\end{document}